\pdfoutput=1
\documentclass[10pt,conference]{IEEEtran}
\IEEEoverridecommandlockouts
\usepackage{cite}
\usepackage{amsmath,amssymb,amsfonts}
\usepackage{graphicx}
\usepackage{textcomp}
\usepackage{xcolor}
\usepackage{float}
\usepackage{stfloats}
\usepackage[caption=false]{subfig}
\usepackage[ruled]{algorithm2e}
\usepackage{algpseudocode}
\graphicspath{{pictures/}}
\usepackage{amsfonts}
\usepackage{cancel}

\def\BibTeX{{\rm B\kern-.05em{\sc i\kern-.025em b}\kern-.08em
    T\kern-.1667em\lower.7ex\hbox{E}\kern-.125emX}}
\begin{document}
\title{Paper Title*\\
{\footnotesize \textsuperscript{*}Note: Sub-titles are not captured in Xplore and
should not be used}
\thanks{Identify applicable funding agency here. If none, delete this.}
}

\newcommand{\etal}{\textit{et al}.}
\newcommand{\ie}{\textit{i}.\textit{e}.}
\newcommand{\eg}{\textit{e}.\textit{g}.}

\def\systemname{\textit{COOK}\xspace}
\def\systemnamenobf{COOK}

\title{\systemnamenobf: Chirp-OOK Communication with Self-reliant Bitrate Adaptation in Backscatter Networks}

\author{\IEEEauthorblockN{Gang Huang\IEEEauthorrefmark{1}, Panlong Yang\IEEEauthorrefmark{1}, Xin He\IEEEauthorrefmark{1}, Yubo Yan\IEEEauthorrefmark{1}, Hao Zhou\IEEEauthorrefmark{1}, Xiangyang Li\IEEEauthorrefmark{1} and Pengjun Wan\IEEEauthorrefmark{2}}
\IEEEauthorblockA{\IEEEauthorrefmark{1}LINKE Lab, School of Computer Science and Technology, University of Science and Technology of China\\ \IEEEauthorrefmark{2} Department of Computer Science, Illinois Institute of Technology, USA\\
Email: {hg2018@mail.ustc.edu.cn, \{plyang, xhe076, yuboyan, kitewind, xiangyangli\}@ustc.edu.cn}}\,wan@cs.iit.edu}

\maketitle

\begin{abstract}
For large-scale Internet of Things (IoT), backscatter communication is a promising technology to reduce power consumption and simplify deployment. However, backscatter communication lacks stability, along with limited communication range within a few meters. 
Due to the limited computation ability of backscatter tags, it is burdensome to effectively adapt the bitrate for the time-varying channel. Thus, backscatter tags are failed to fully utilize the optimal transmission rate.
In this paper, we design a system named COOK with self-reliant bitrate adaptation in backscatter communication. 
Channel symmetry allows backscatter tags to adjust bitrate depending on the received signal strength of the excitation source (ES) without feedback.
In addition, the chirp spreading signal is exploited as the ES signal to enable backscatter tags to work under noise floor.
Our modulation approach is denoted as Chirp-OOK since the tags reflect the chirp signal by employing the on-off keying modulation. It allows that receiver can decode under the noise floor and the bitrate varies flexibly as the communication range changes.
We have implemented the prototype system based on the universal software radio peripheral (USRP) platform. Extensive experiment results demonstrate the effectiveness of the proposed system.
Our system provides valid communication distance up to $27m$, which is $7\times$ as compared with normal backscatter system.
The system significantly increases the backscatter communication stability, by supporting bitrate adaptation ranges from $0.33kbps$ to $1.2Mbps$, and guaranteeing the bit error rate (BER) is below $1\%$.

\end{abstract}

\section{Introduction}
\label{sec:intro}
With the rapid development of the Internet of Things industry, the number of network devices has increased exponentially. Connecting a large number of devices is a challenging task. The wired connection usually is not a feasible solution, and wireless connection is still energy-hungry and costly. 
Fortunately, backscatter communication~\cite{liu2013ambient,kellogg2015wi,wang2017fm,talla2017lora,bharadia2015backfi,peng2018plora} makes wireless transmissions at a power consumption orders of magnitude lower than traditional radios. 
The reason is that backscatter communication reflects ES signals instead of generating the RF signal itself. Due to the ultra-low power consumption feature and easy deployment, backscatter communications become a good solution to connect IoT devices. 

A new generation of backscatter tags can be equipped with a variety of sensing functions that communicate communications. They can also be equipped with a small on-board battery for sensing and calculation. Recent studies indicated that industries of health care, retail, oil, and aerospace were moving towards object tracking, asset monitoring, and machine-to-machine (M2M) applications, leading to deployment such backscatter networks. 

Recently, the challenges of using backscatter communication are the relatively short communication \textbf{range} and lower \textbf{bitrate}. Extensive researches have been proposed to solve these two challenges.
(1) Peng \etal~\cite{peng2018plora} proposed PLoRa, a passive LoRa communication technology that enables long-distance connectivity of IoT devices. However, LoRa-based systems employed a chirp spread spectrum (CSS) modulation scheme, causing very low bitrate 18 bps.
(2) Dinesh \etal~\cite{bharadia2015backfi} proposed BackFi, providing at least $1$-$5Mbps$ of uplink throughput with only $5$ meters of maximal range.
From the above-mentioned studies, the communication rate and range were not improved at the same time. Consequently, backscatter networks require adaptive data rate (ADR) to balance the communication range and rate, especially in time-varying channels. In this paper, we investigate the problem of ADR in backscatter networks. Essentially, the major challenges of this problem are as follows:

\begin{itemize}
  \item
  The tags must continuously estimate the channel quality by probing~\cite{bicket2005bit,judd2008efficient} or by requiring channel state feedback~\cite{wang2012efficient,gudipati2011strider,vutukuru2009cross} from the receiver. However, this feedback is not effective and will result in bitrate adaptive delays. On the other hand, the receiver-to-tag link may not even exist in some scenario due to the simplicity of the tags. 
  This means that it is very difficult to obtain feedback on the channel state of the receiver for ADR.
  \item
    The wireless signal will attenuate during the propagation process, and the backscatter signal will be more severely attenuated due to reflection. And we can't simply increase the ES transmission power, which is unsafe and inefficient. For existing backscatter systems with tags more than $2m$ away from the ES, the backscatter signal will be submerged under the noise floor eventually.
\end{itemize}

In order to solve the aforementioned problems, we design a system named \systemname that balances the communication range and rate, even if the signal to noise ratio (SNR) of backscatter signal is less than $0 dB$. 
In Table~\ref{TABLE:Comparison}, we compared our system with existing backscatter systems. And our contributions are based on the following design elements:

\begin{itemize}
  \item
  \textit{Feedback-free bitrate adaptation.}
  We verify the channel symmetry, then draw to conclusion that the SNR at the receiver is related to ES signal strength at the tags. This means that the tags can be bitrate-adaptive and  only requires to know the ES signal strength.
  \item
  \textit{Doable under the noise floor.}
   We achieve backscatter communication under the noise floor by importing chirp signal. Unlike other chirp signal based backscatter schemes, only the ES produces chirp signal instead of tags.
   The advantage is that the tags still use the original on-off keying (OOK) modulation, and flexibly change the OOK square wave unit length for rate adaptation. The cost of the tags will not change a lot.
  \item
  \textit{Flexible decode and frame mechanism.}
  We design of the frame during modulation in tags, and design a dynamic threshold algorithm during demodulation. Then, the BER of \systemname is reduced to below $1\%$ when the backscatter signal is below the noise floor. 
\end{itemize}

\vspace{-0.1in}
\begin{table}[ht]
	\centering
	\scriptsize
	\footnotesize
	{\caption{Comparison of COOK with existing backscatter systems}
		\label{TABLE:Comparison}
	}
	\vspace{-0.1in}
	{
		\begin{tabular}{c|ccc}
			\hline
			\bfseries Technology & \bfseries Data Rates & \bfseries ES-tag distance  \\
			\hline
			Ambient Backscatter~\cite{liu2013ambient} & 1kbps & $\leq$1m \\
			WiFi Backscatter~\cite{kellogg2015wi} & 1kbps & 0.65m  \\
			FM Backscatter~\cite{wang2017fm} & 3.2kbps & $\leq$18m  \\
			Lora Backscatter~\cite{talla2017lora} & 8.7bps & $\leq$237.5m \\
			BackFi~\cite{bharadia2015backfi} & 10kbps-5Mbps & 5m  \\
			PLora~\cite{peng2018plora} & 6.25kbps & $\leq$1m \\
			\textbf{COOK} & \textbf{0.33kbps-1.2Mbps} & \textbf{$\leq$27m}  \\
			\hline
		\end{tabular}
	} 

\end{table}

The rest of this paper is organized as follows. Sec.~\ref{SEC:Primer} introduces backscatter network and CSS. 
Sec.~\ref{SEC:Model} will detail the model of channel symmetry and Chirp-OOK. Sec.~\ref{SEC:System} will detail the mechanism of tags and receiver. Sec.~\ref{SEC:Implementation} is system implementation and evaluation followed by the conclusion of the paper.
\vspace{-0.05in}

\section{Preliminary and Background}
\label{SEC:Primer}

\begin{figure*}[htb]
\centering
\begin{minipage}[t]{0.35\linewidth}
     \centering    
     \includegraphics[width=0.9\columnwidth,height=3cm]{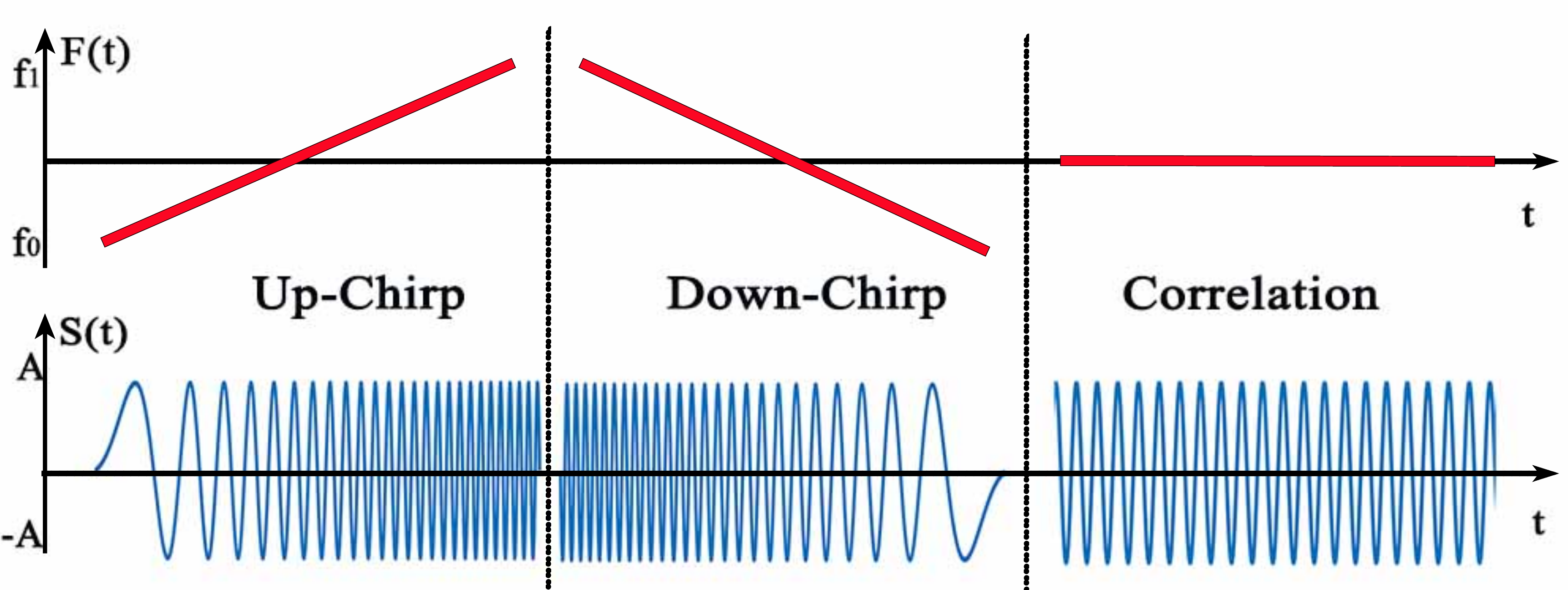}
    \caption{Up-chirp, down-chirp and correlation signal in time domain and frequency domain}
    \label{FIG:CSS}
\end{minipage}
\begin{minipage}[t]{0.24\linewidth}
    \centering
    \subfloat[monostatic]{
        \label{SUBFIG:deployment2}
        \includegraphics[width=0.45\linewidth,height=3cm]{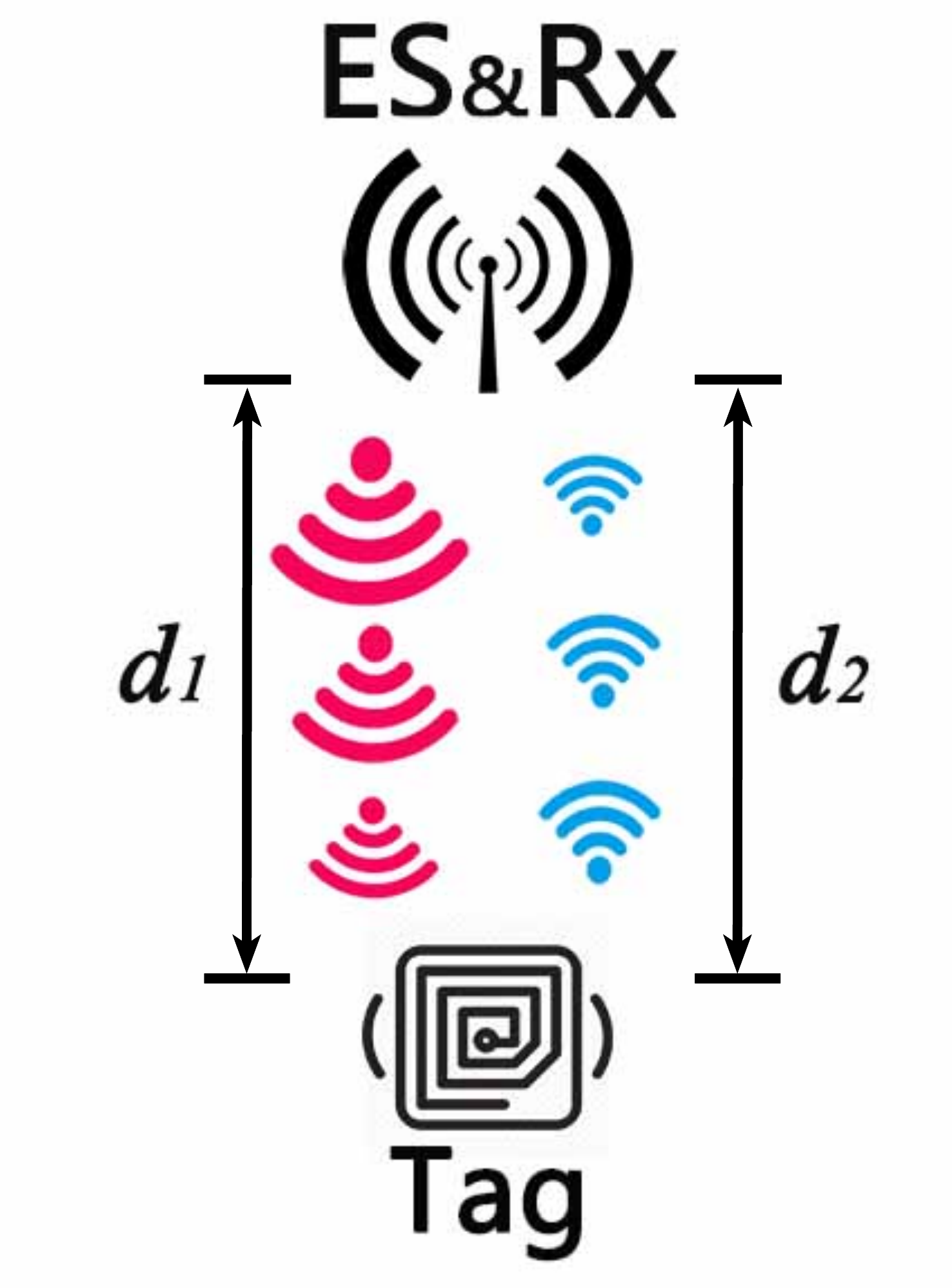}
        }    
	\subfloat[bistatic]{
	    \label{SUBFIG:deployment1}
	    \includegraphics[width=0.45\linewidth,height=3cm]{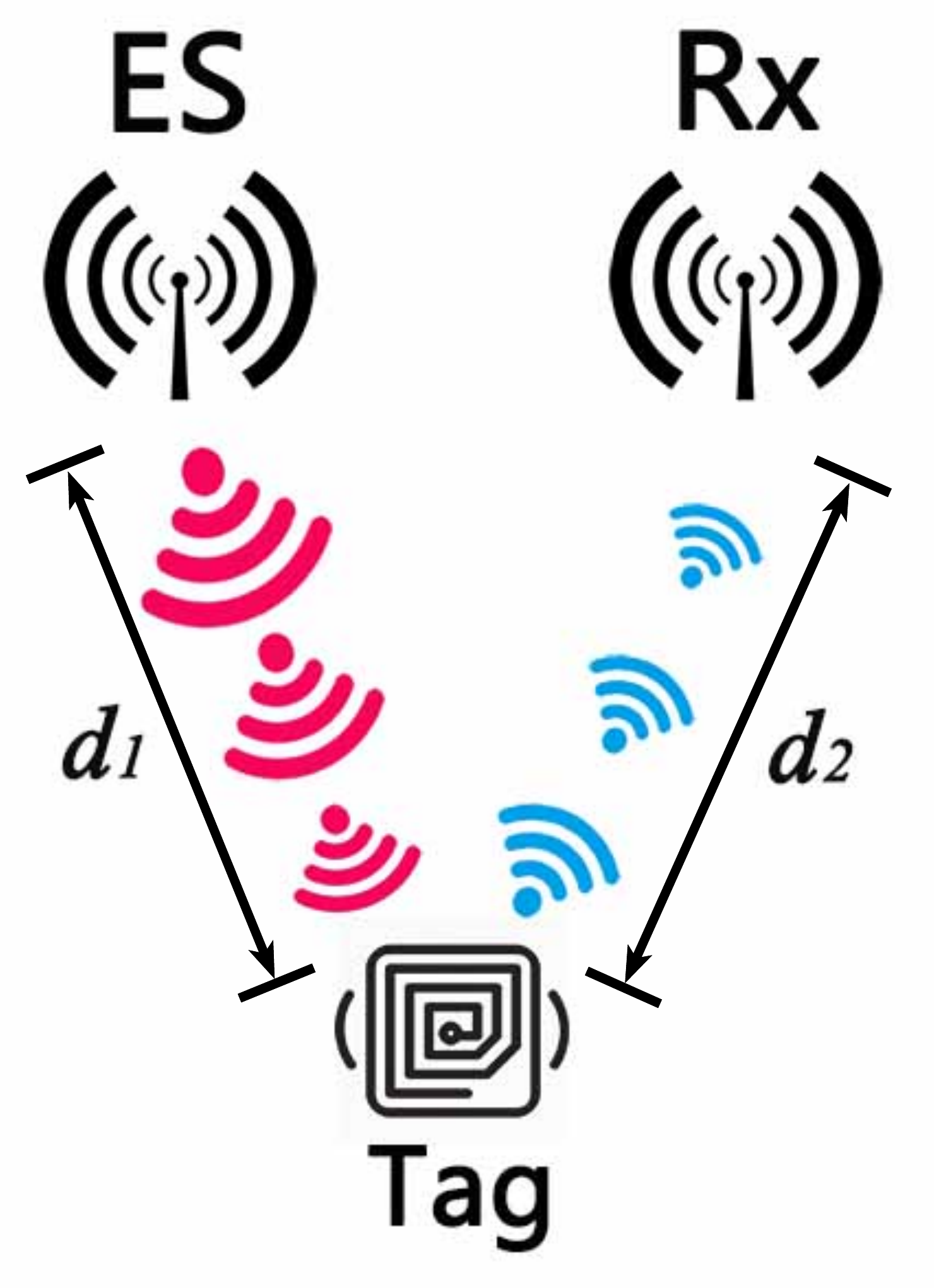}
	    }
	\caption{Backscatter system deployments }
		    \vspace{-0.1in}
    \label{FIG:deployment}
\end{minipage}
\begin{minipage}[t]{0.35\linewidth}
	\centering
	\includegraphics[width=0.9\columnwidth,height=3cm]{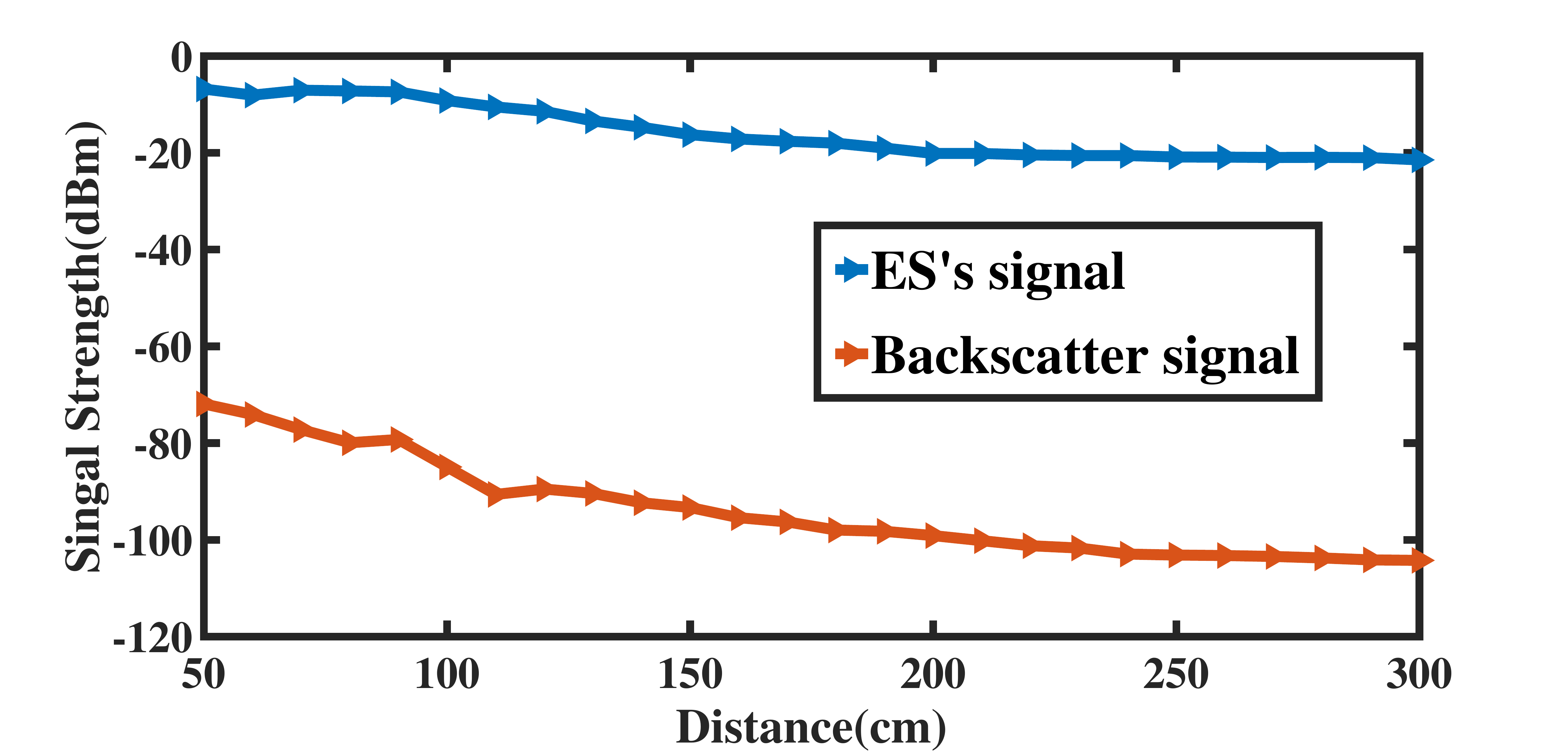}
	\caption{Two signal strength with varied distance }
		\vspace{-0.1in}
	\label{FIG:duichen}
\end{minipage}
\end{figure*}

\subsection{Backscatter Communication}
\label{SEC:Backscatter_primer}
In a backscatter network, the excitation source transmits a high power continuous waveform. The backscatter tag transmits its signal by reflecting the continuous waveform using on-off keying. The bit '1' is then transmitted by changing the impedance on its antenna to reflect the ES signal and to define the bit '0' by maintaining the initial state~\cite{camp2010modulation}. There are three main differences between the backscatter network and the more familiar wireless network:
(1) The tags do not generate their own RF signals but reflect the ES signal, every backscatter tags should need more than one ES.
(2) Backscatter tags transmit within a narrow bandwidth due to their power limitations. Therefore, backscatter communication is sensitive to time-varying channel.
(3) The tags transmit data through the excitation and cannot perform complex decoding operations due to their power limitations. But receiver is power-supplied and can delegate decoding complexity to it, while keeping the backscatter tags simple and energy efficient.
\vspace{-0.05in}
\subsection{Chirp Spread Spectrum}
\label{SEC:CSS_primer}
In digital communications, chirp spread spectrum (CSS)~\cite{berni1973utility} is a spread spectrum technique that uses wideband linear frequency modulated chirp pulses to encode information. In current commercial communication systems, CSS communication is the only communication method that can operate under negative SNR conditions. Suppose the duration of the chirp pulse is $T$, within this time domain the frequency increases (up-chirp) or decreases (down-chirp) monotonically. As shown in Fig.~\ref{FIG:CSS}, the difference between up-chirp and down-chirp frequencies can approximate the chirp pulse’s bandwidth, $B$.  We can express up-chirp and down-chirp as formula 
\begin{align}
\begin{array}{ll}{\text { up-chirp: }} & {S_{1}=A \cdot e^{j\left(2 \pi f_{0} t+\pi \cdot k / 2 \cdot t^{2}\right)}} \\ {\text { down-chirp: }} & {S_{2}=A \cdot e^{j\left(2 \pi f_{1} t-\pi \cdot k / 2 \cdot t^{2}\right)}}\end{array}.
\label{EQN:upchirp-and-downchirp}
\end{align}
The $A$ is signal amplitude, $f_{0}$ and $f_{1}$ state for the initial frequency of up-chirp and down-chirp respectively, $k$ is roll-off rate of frequency. We can multiply up-chirp by down-chirp to get the correlation signal at the receiver:
\begin{align}
S = A \cdot e^{j\left[2 \pi (f_{0}+f_{1}) t\right]}.
\label{EQN:chirp}
\vspace{-0.1in}
\end{align}

After the FFT (Fast Fourier transforming) the correlation signal, the correlation signal peak value in the spectrogram will depend on the sampling time, which is the duration $T$ of the chirp signal.

\section{Model}
\label{SEC:Model}

\subsection{Channel Symmetry}
\label{sec:basis-of-Channel}

In the absence of feedback, channel symmetry allows the backscatter tags to adjust bitrate by estimating the signal strength at the tag. We will introduce channel symmetry in two deployment scenarios below.

As shown in Fig.~\ref{FIG:deployment}, there are two different deployment scenarios in the backscatter network. 
One is called monostatic, which puts the ES together with the receiver. The other one is called bistatic, where the ES is separated from the receiver 
Consider the backscatter signal of strength $P_{r}$ in the receiver in free space, 

\begin{align}
    P_{r}=\Gamma \cdot \left(\frac{P_{t} G_{t}}{4 \pi d_{1}^{2}}\right) \cdot \left(\frac{\lambda^{2} G_{r}}{4 \pi d_{2}^{2} 4 \pi}\right).
    \label{model}
\end{align}
Here, $\lambda$ is the carrier's wavelength, $P_{t}$ is ES transmission power, the factor $\Gamma$ is constant ratio between the return loss and antenna gained at the backscatter tags. $G_{t}$ and $G_{r}$ represent the antenna gain for ES and receiver. Similarly, $d_{1}$ denotes the distance from the backscatter tags to the ES and $d_{2}$ denotes the distance from the tags to the receiver. 

\emph{Monostatic.} In the monostatic scenario, the ES and the receiver are placed in one device, which greatly reduces deployment costs. This design of using one device containing both ES and receiver ensures the symmetry of the up-link and down-link channel, i.e., $d_{1} = d_{2}$ and $G_{t} = G_{r}$. If ES transmission power $P_{t}$ is already known, the formula for $P_{r}$ can be simplified as 
\vspace{-0.05in}
\begin{align}
    P_{r}= \Gamma \cdot \frac{\lambda^{2}}{(4 \pi)^{3}} \cdot P_{t} \cdot  (\frac{G_t}{d_{1}^{2}})^2.
    \label{EQN:model1}
\end{align}
When tags receiving ES signal, the signal strength $P$ at the tags could be regard as $\frac{P_{t} G_{t}}{4 \pi d_{1}^{2}}$. Therefore, we can estimate $P_r$ based on $P$ as follow. By simple transformation of Eq.~\ref{EQN:model1},
\vspace{-0.05in}
\begin{align}
    P_{r}=\Gamma  \cdot \frac{\lambda^{2}}{4 \pi} \cdot P_{t} \cdot P^{2}=\eta \cdot P^{2}.
    \label{EQN:model3}
\end{align}
Here, $\eta = \Gamma \cdot \frac{\lambda^{2}}{4 \pi} \cdot P_{t} $ is a constant value which is independent to positions of tags, ES and receiver. And Eq.~\ref{EQN:model3} suggests that tags can adjust the bitrate by estimating $P$ instead of $P_r$.  

The channel symmetry is also verify in our experiments. We place the ES with receiver and moving the tags away from them, then we collected the ES signal strength at the tags and the backscatter signal strength at the receiver. We collect long-term data in different environments, and the result is shown in the Fig.~\ref{FIG:duichen}.  We notice that the strength of two signals changed in parallel, and two signals strengths fall in the same trend as the distance increases.

\emph{Bistatic.} The advantage of bistatic scenario is that the tags can be placed near the ES so that the receiver can be far apart, i.e. $d_1 < d_2$,  which is suitable for outdoor environments.
As shown in Fig~\ref{SUBFIG:strength2}, the intensity attenuation near the ES in bistatic scheme is similar in monostatic scheme. So when the tag is near the ES ($d_1 < d_2$),  Eq.~\eqref{EQN:model1} is still applicable.
\begin{align}
    P_r= \eta\cdot P^2, \quad (d_1 < d_2).
    \label{EQN:model4}
\end{align}

In summary, whether in monostatic or bistatic ($d_1 < d_2$) scenario, the tags can estimate the ES signal strength $P$ for bitrate adaptation based on channel symmetry.

\subsection{Chirp-OOK}
\label{sec:chirp-ook}

\begin{figure}[!ht]
    \vspace{-0.1in}
    \centering
	\subfloat[\footnotesize Monostatic]{
	    \label{SUBFIG:strength1}
	    \includegraphics[width=0.23\textwidth,height=3cm]{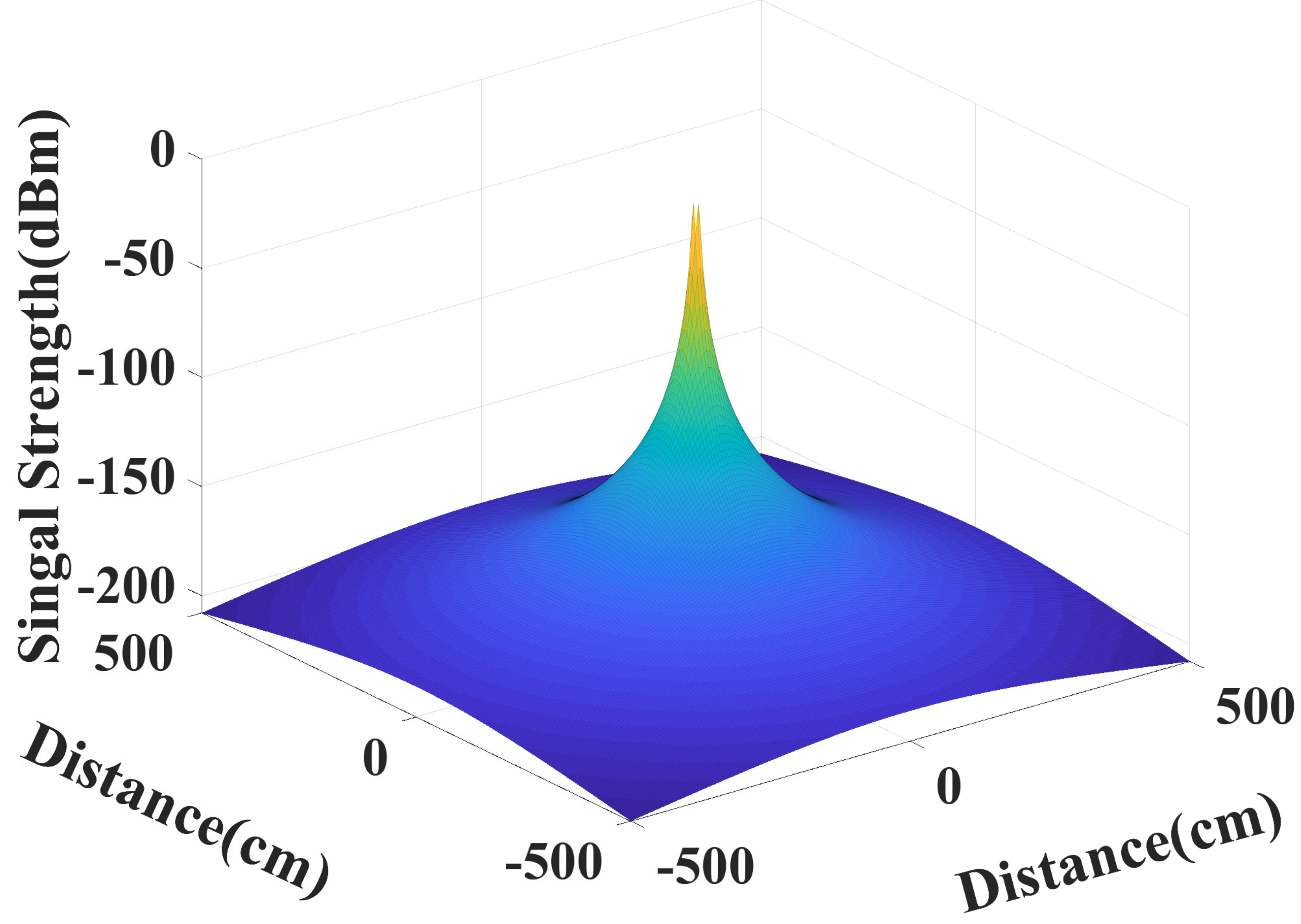}}
	\subfloat[\footnotesize Bistatic]{
	    \label{SUBFIG:strength2}
		\includegraphics[width=0.23\textwidth,height=3cm]{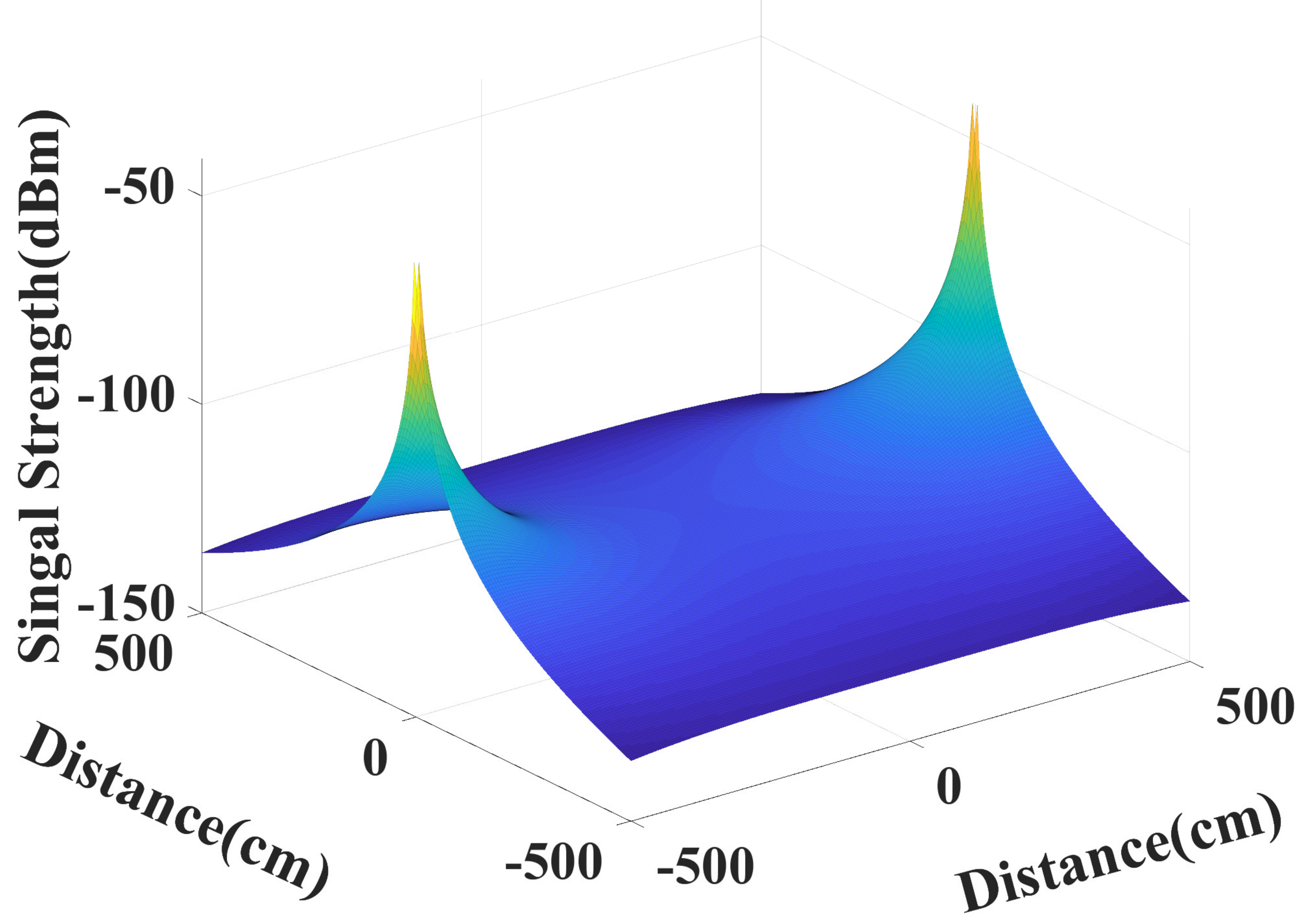}}
	\caption{Distribution of signal strength.}
	    \vspace{-0.1in}
    \label{FIG:strength}
\end{figure}
As shown in the Fig.~\ref{FIG:strength}, the backscatter signal is weak when the tags is far away from the ES in either monostatic or bistatic.  
Our experiment results suggest that the signal strength is already below $-120dbm$ when ES-Tag distance is more than $2m$.
Thus the effective working range of this program is limited.

In order to enable receiver successful decoding weak tags signal even lower than noise floor, we use chirp signal as ES signal and adapt CSS modulation scheme.
Unlike traditional CSS scheme that the ES genebitrates tone signal and tags produce chirp signal. In contrast, the tags remain on-off keying. Only ES recursively generates up-chirp signal, then receiver uses a down-chirp of opposite roll-off rate to correlate up-chirp signal. From Eq.~\eqref{EQN:chirp}, we know the correlated signal frequency is a fixed value. So we can perform FFT to determine whether there a backscatter signal based on the peak value on the FFT bin. For FFT, we have the following:
\begin{align}
f_{k}=\sum_{n=0}^{N-1} t_{n} \cdot e^{-i 2 \pi k n / N},
\label{EQN:FFT}
\end{align}
where $f_{k}$, and $t_{n}$ represent values in the frequency domain and the time domain, and $N$ is the number of sampling points for chirp signal. The $f_{k}$ will be accumulated with the same frequency, and the peak value is the largest $f_{k}$,
\begin{align}
Peak=\operatorname{Max}\left\{f_{k}\right\},k \in [1,N].
\label{EQN:Peak}
\end{align}
The $ Peak \propto N $, the $Peak$ value will increase as the number of sampling points grows. 
We compare the value of peak with the threshold to determine whether the bit is '1' or '0'. 
The threshold settings are detailed in Sec.~\ref{SEC:receiver}. 
And this modulation method combines chirp signal with on-off keying, so we call it Chirp-OOK.
The backscatter tags can change the size of the $Peak$ value by adjusting the unit length of the on-off square wave.

\section{System Design}
\label{SEC:System}

\begin{figure}[htbp]
	\centering
	\includegraphics[width=0.45\textwidth]{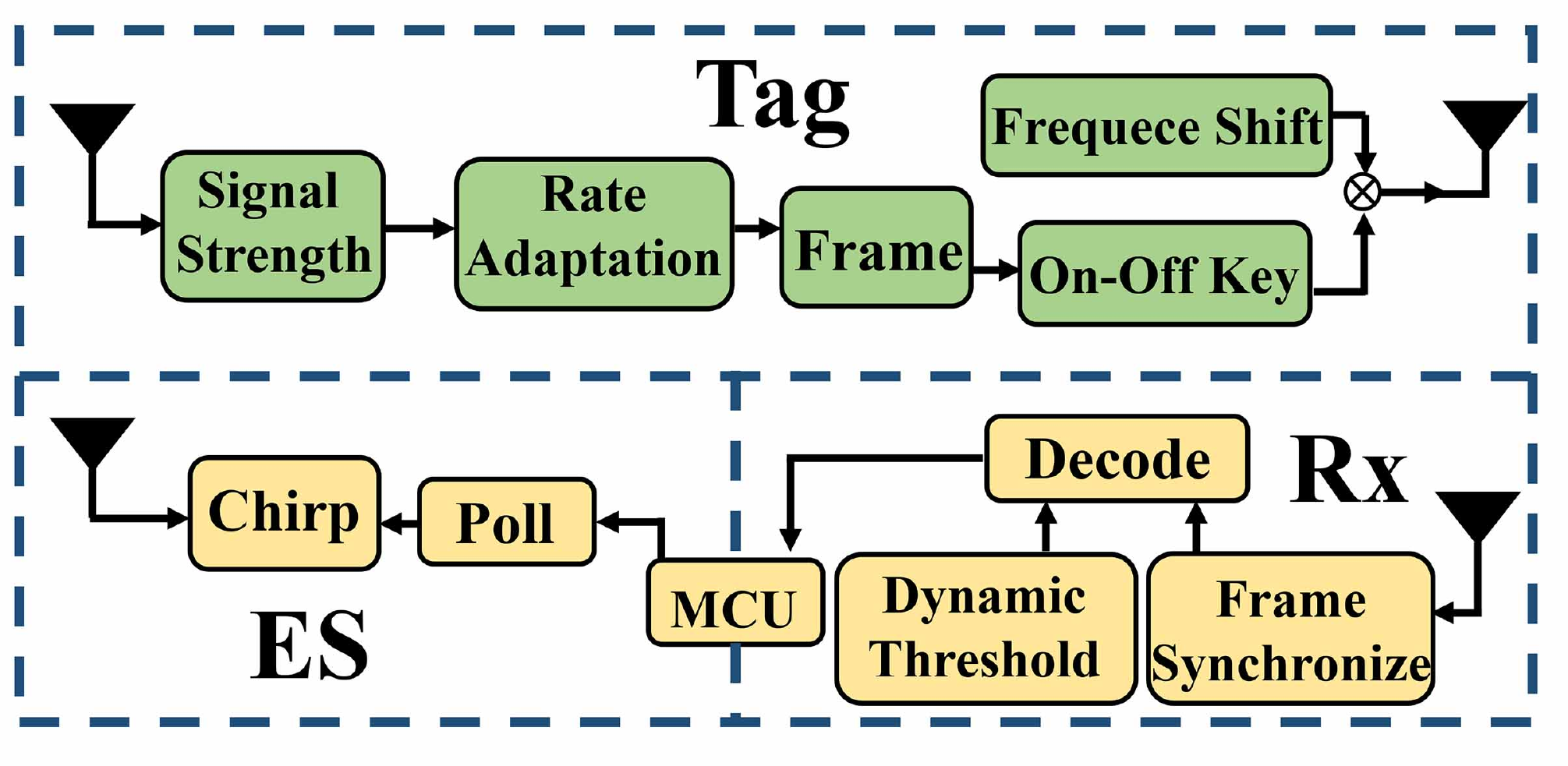}
	\vspace{-0.1in}
	\caption{System workflow}
	\label{FIG:workflow}
\end{figure}

The COOK system consists of three parts containing ES, receiver and several tags. The ES broadcasts a cyclic chirp signal, also serves as an energy source for the tags. The tags passively transmits data frames to the receiver by reflecting and modulating the ES signal. Tags actively adjusts bitrate based on detecting the strength of ES signal. The receiver obtains the bitrate by decoding the preamble and performs dynamic threshold algorithm. The receiver uses a polling mechanism for reading multiple tags. The data of different tags are preferentially read according to their bitrate. The structure of the system is shown in Fig.~\ref{FIG:workflow}.
\vspace{-0.1in}

\subsection{Tag}
\label{SEC:tag}

\textbf{Self-reliant bitrate adaptation.}
In order for the receiver to demodulate successfully, the tags needs to adjust the on-off square wave unit length  ($L$)  through the ES signal strength $P$.

In information theory, the shannon-hartley theorem shows the channel capacity and illustrates that the maximal bitrate.
\vspace{-0.05in}
\begin{align}
     C=B  \log _{2}(1+{S}/{N})
    \label{EQN:shannon1}
\end{align}

where the channel bandwidth $B$ is constant, the $S/N$ = signal-to-noise ratio (SNR) and $C$ (channel capacity) of our system needs to be adapted.
When signal strength is lower than noise floor, i.e., $S/N << 1$. So
\vspace{-0.05in}
\begin{align}
        C= \frac{B}{\ln 2} \times \ln (1+S / N) \approx 1.433\times B \times {S} / {N}.
	\label{EQN:shannon1-1}
\end{align}

As shown in Eq.~\eqref{EQN:model3}, the signal strength $S$ in our system is  $S = P_r = \eta  P^2$, which is in $dbm$ term. By substituting Eq.~\eqref{EQN:model3} into Eq.~\eqref{EQN:shannon1-1}, we have
\begin{align}
	 C=1.433 \cdot B \cdot 10^{\sqrt{\frac{\eta}{10}} \cdot 2 P} / N.
	\label{EQN:shannon3}
\end{align}
By measuring in advance, we can get the maximum bitrate $C_0$ and the corresponding strength of the signal $P_0$. Finally, 
\vspace{-0.05in}
\begin{align}
	 C = C_0\cdot10^{K(P-P_0)}.
	\label{EQN:rate}
\end{align}
where $K=2\sqrt{\frac{ \eta }{10}}$ represents the return loss and antenna gains at the backscatter tags in communication process. 
$K$ only related to carrier’s  wavelength $\lambda$, ES transmission power $P_t$ and tags loss ratio $\Gamma$. 
The $K$ also can be call environment-independent factor, which could be measured and set to tags in advance. 

As far now, we can conclude that $L=\frac{1}{C} = \frac{1}{C_0\cdot10^{K(P-P_0)}}$ could guarantee receiver successful decoding.

\begin{figure}[htbp]
    \centering
        \centering
        \includegraphics[width=0.40\textwidth]{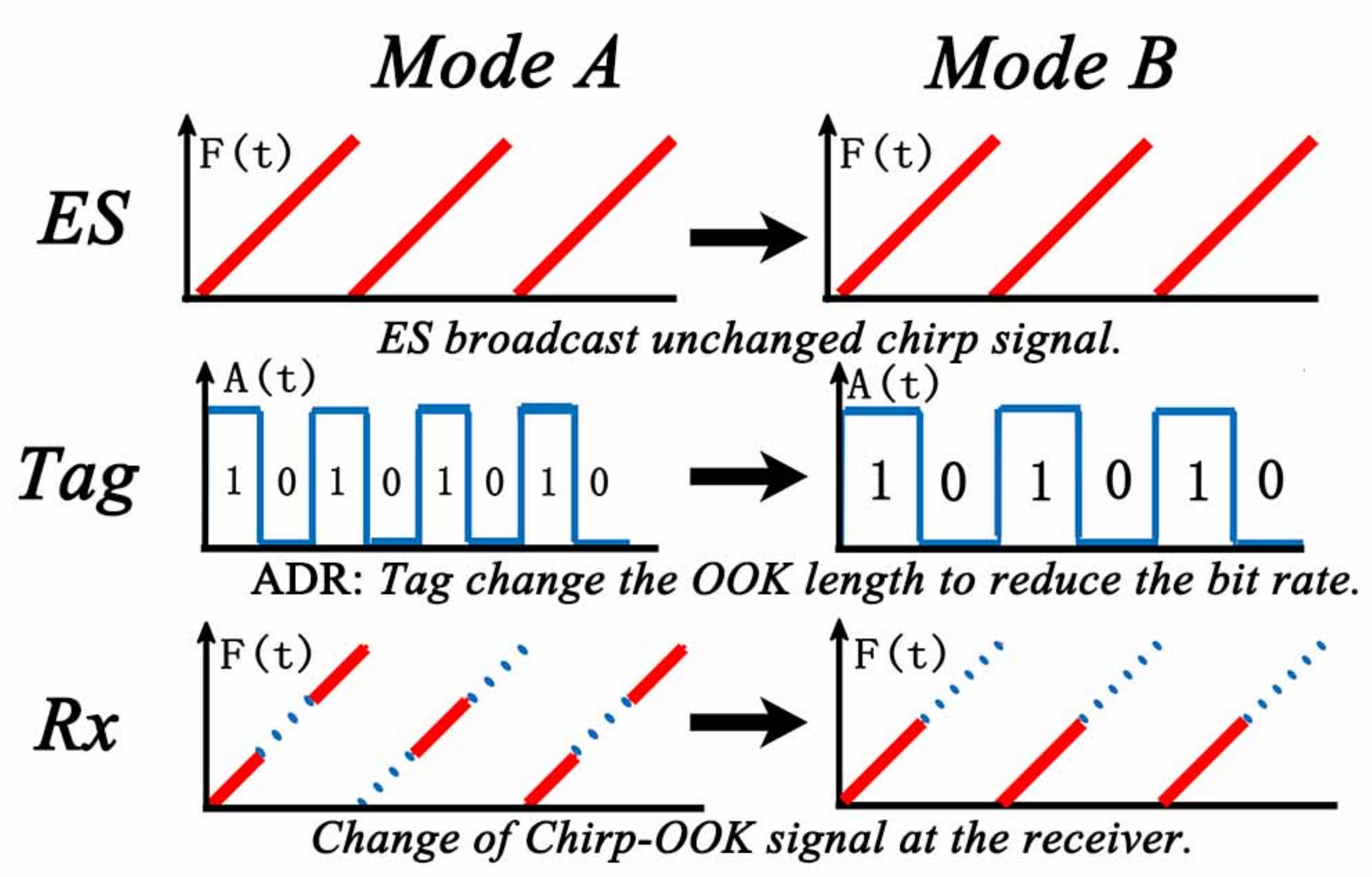}
        \caption{Bitrate adaptation process}
        \label{FIG:rate_adjust}
        \vspace{-0.1in}
\end{figure}
An example of bitrate adaptation is shown in the Fig.~\ref{FIG:rate_adjust}. When the SNR decreases, the tags will decrease in bitrate from Mode A to B. We can see that in this process the tags only need to change the unit length of the on-off square wave for automatic bitrate adaptation. The ES signal remains unchanged. It greatly reduces the complexity of the system.

\textbf{Framing.}
The  receiver  needs  to  synchronize  and  should  know the  unit length   of  on-off  square  wave. 
Otherwise, the FFT peak value cannot distinguish between bit '1' and bit '0' when the backscatter signal is under the noise floor.

\begin{figure}[htbp]
    \centering
	\includegraphics[width=0.3\textwidth]{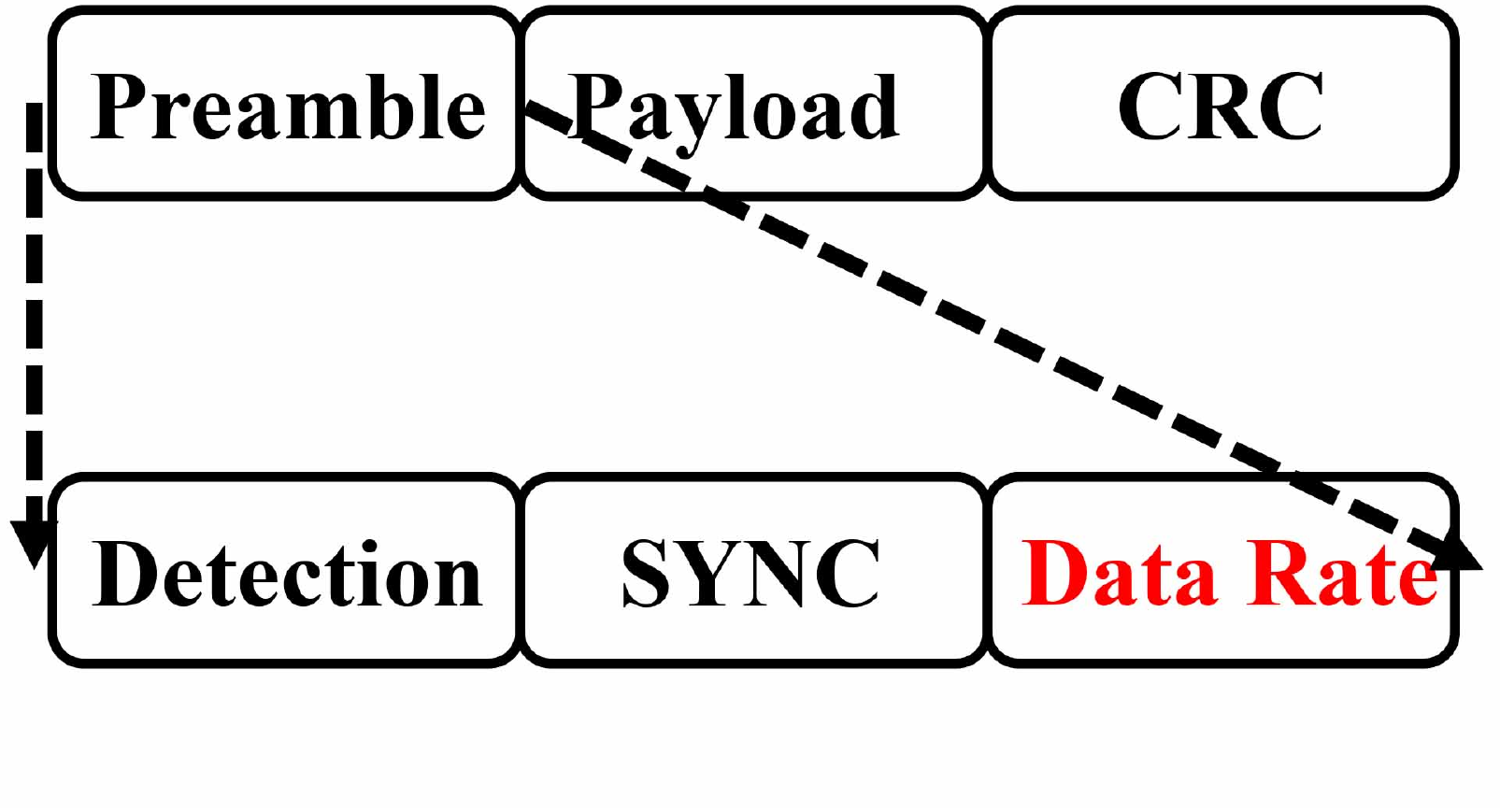}
	\vspace{-0.2in}
	\caption{Frame structure}
	\label{FIG:phy_frame}
\vspace{-0.1in}
\end{figure}

We designed a physical layer frame structure, as shown in Fig.~\ref{FIG:phy_frame}. Before the payload, we added the detection field, sync field and the data rate field as preamble, and put a cyclic redundancy check (CRC) at the end of the frame to ensure the correctness of the transmitted data. In the detection field and sync field, the on-off square wave unit length is fixed and takes the maximum which is known for receiver.
Data rate field can be calculated as on-off square ware unit length of payload.
In this way, the receiver can obtain the bitrate of the payload data by decoding the preamble.

\vspace{-0.1in}

\textbf{Frequency shift}. 
Our system supports multiple tags. In order to avoid interference among tags, we adopt the frequency shift technology~\cite{zhang2016hitchhike}. The system could receive signal from multiple tags at different center frequency. This improves the concurrency of the system.

\subsection{Receiver}
\label{SEC:receiver}

\textbf{Decode.}
The decode process consists of two parts, one is a sliding window algorithm for synchronization, and the other is a dynamic threshold algorithm for demodulation.

Receiver uses the sliding window algorithm to detect if up-chirp signal is coming, then align it with the down-chirp signal. The down-chirp signal length is the same as the on-off square wave unit length in the preamble. If the up-chirp signal is aligned with the down-chirp signal, the FFT peak is the largest. Therefore, think of down-chirp as a window, and the window slides until the end of the maximum FFT peak value, then the alignment process is complete.

After the alignment, receiver can get the bitrate of payload by decoding the preamble, which represents the on-off square wave unit length of payload. 
In the traditional OOK demodulation method, a fixed threshold is used to determine whether the signal amplitude is '1' or '0'. 
When there are unpredictable channel changes, the FFT peak value is likely to radical change in one data frame, especially under noise floor. Traditional OOK demodulation method is unworkable when the threshold is fixed. 
In our system, we used a dynamic threshold algorithm to solve this problem. Algorithm.~\ref{alg:Dynamic} is our dynamic threshold algorithm, which can update the threshold in real time by using the past FFT peak value. 
To get the threshold used in determining the $i_{th}$ peak $Peak_i$, we calculate the ratio $p=\left(P e a k_{i-1}-P e a k_{i-2}\right) / P e a k_{i}$ , 
then the updated threshold is $thr_{i}  = thr_{i-1} \cdot (1 + p)$.
Here the initial threshold is usually set to be $Peak_1 /2$, where $Peak_1$ is first peak value.
Now if the FFT peak value is greater than the current threshold, the bit is '1' and opposite is '0'.

\vspace{-0.1in}
\begin{algorithm}[thb]
	\caption{Dynamic Threshold Decode}
	 \label{alg:Dynamic}
	\LinesNumbered 
	\KwIn{signal $R$, down-chirp $D$, chirp length $l$,code numbers $n$}  
	\KwOut{chips}
	 $Peak \gets [\,]$,$chip \gets [\,]$\;
	 $Data \gets S(R,start,l)$ get a segment of the received signal.\;
	 $Peak_{1,2} \gets P(Data,Down$-$Chirp)$ get FFT peak.\;
	 $chip_{1,2} \gets 1$;
	 $threshold \gets Peak_{1}/2$\;
	\For {$i = \  3 \ \to \ n$ }{
		 $data \gets S(R,start,L)$\;
		 $peak \gets P(data,D)$\;
    	\If{$Peak > threshold $}{
    	    $threshold \gets threshold*(1+\frac{Peak_{i-1}-Peak_{i-2}}{Peak_{i}})$
    		$chip_{i} \gets 1$\\
	\Else{
	    $chip_{i} \gets 0$
	    }
	}
    	$start \gets start + l$;
        $i \gets i+1$
    }
\end{algorithm}

\textbf{Polling.} 
Consider multiple tags in system, the receiver uses the Listen Before Talk (LBT) mechanism and allocates read priority for different tags.  

The receiver adjusts the center frequency of the ES one by one according to the query list to detect whether the tag exists. If the data frame of the tag is accepted, the tag is inferred to be within the read range, otherwise the tag is considered to be out of range. Then ES will poll the results in the table and update the table every fixed time. Our tags are passive and ES can be turned off when the receiver does not need to read the tag's data. The advantage of this mechanism is that ES don't need to stay open in long time for some low duty cycle tags, which greatly reduces the power consumption of the entire system.

With multiple tags and only one receiver, the read priority need to be allocated reasonably if different tags need be read at same time. Receiver assign the priority to the tag based on the bitrate value of different tags. We guarantee that the highest bit rate tag occupies the highest priority, since a higher bit rate means that transmission channel of the tag has a larger SNR. When the decoding fails, we re-listen the channel until it is successfully decoded.

\section{System Implementation and Evaluation} 
\label{SEC:Implementation}
\begin{figure}[htbp]
	\centering
	\includegraphics[width=0.3\textwidth]{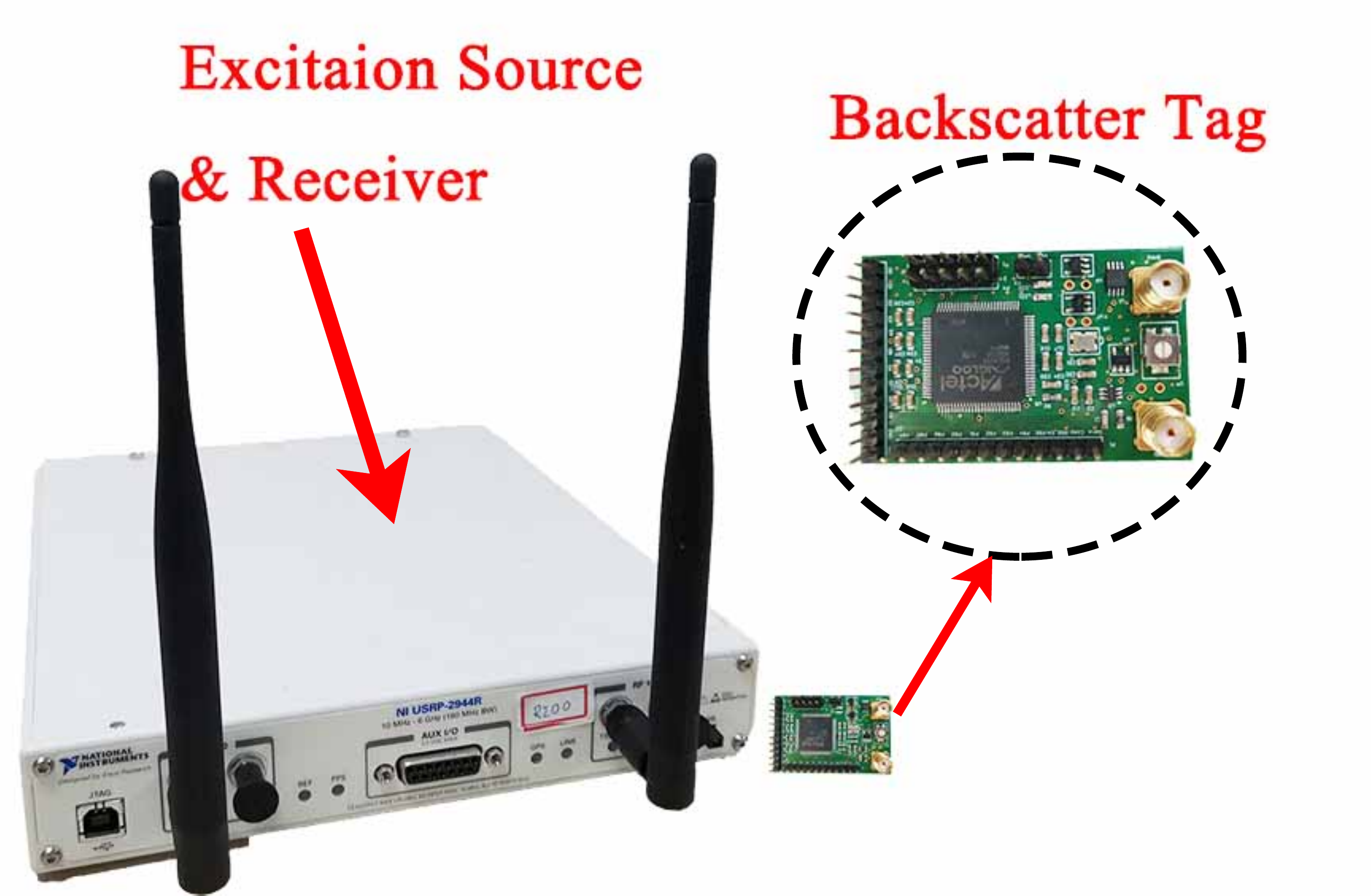}
	\caption{Prototype of \textbf{\systemname}}
		\vspace{-0.1in}
	\label{FIG:impl}
\end{figure}
\vspace{-0.05in}
\subsection{System Prototype}
\label{System Prototype}
Our receiver and ES are implemented on the latest NI USRP RIO platform with the GNU-Radio SDR platform. In order to minimize the interference between the excitation source and the receiver, we put them on two daughter boards. The USRP RIO connects two vertical antennas VERT900 with two (824-960 MHz, 1710-1990 MHz) dualbands as receiving and transmitting antennas. We configure the USRP to broadcast at power levels such that each amplifier outputs $20dBm$ when transmitting in order to stay within FCC limits, and set the carrier frequency of the system to about $900Mhz$ to stay in the ISM band. The prototype of our tags are come from HitchHike~\cite{zhang2016hitchhike} platform. The Fig.~\ref{FIG:impl} depicts our system prototype.

\emph{Excitation Source.}
We use the voltage controlled oscillator (VCO) block in GNU-Radio to loop the ES to generate the chirp signal. We interpolate and sample at [0,1] to get a uniform floating point sequence and output it to the VCO block. It adjusts the resonant frequency based on the input value to produce an up-chirp signal. The VCO block provides two parameters, sampling rate and sensitivity. By changing these two values, we can get up-chirp signal with different bandwidths. On the other hand, at the beginning of the system work, we input the obtained sequence into the VCO block in reverse order, and store the output value of the VCO in the local file of the receiver as the down-chirp signal.

\emph{Receiver.}
The receiver can change the receiving frequency according to the list of frequency shifts of the tags stored in the local file. In addition, the receiver can adjust the excitation sources by the information obtained since they are on the same device.

\emph{Backscatter Tag.}
The important components of the tag are divided into 3 blocks, detection block, reflection block and field programmable gate arrays (FPGA).

\begin{figure}[!ht]
	\centering
    \vspace{-0.15in}
    \subfloat[\footnotesize Different $K$ for BER]{
	\includegraphics[width=0.22\textwidth]{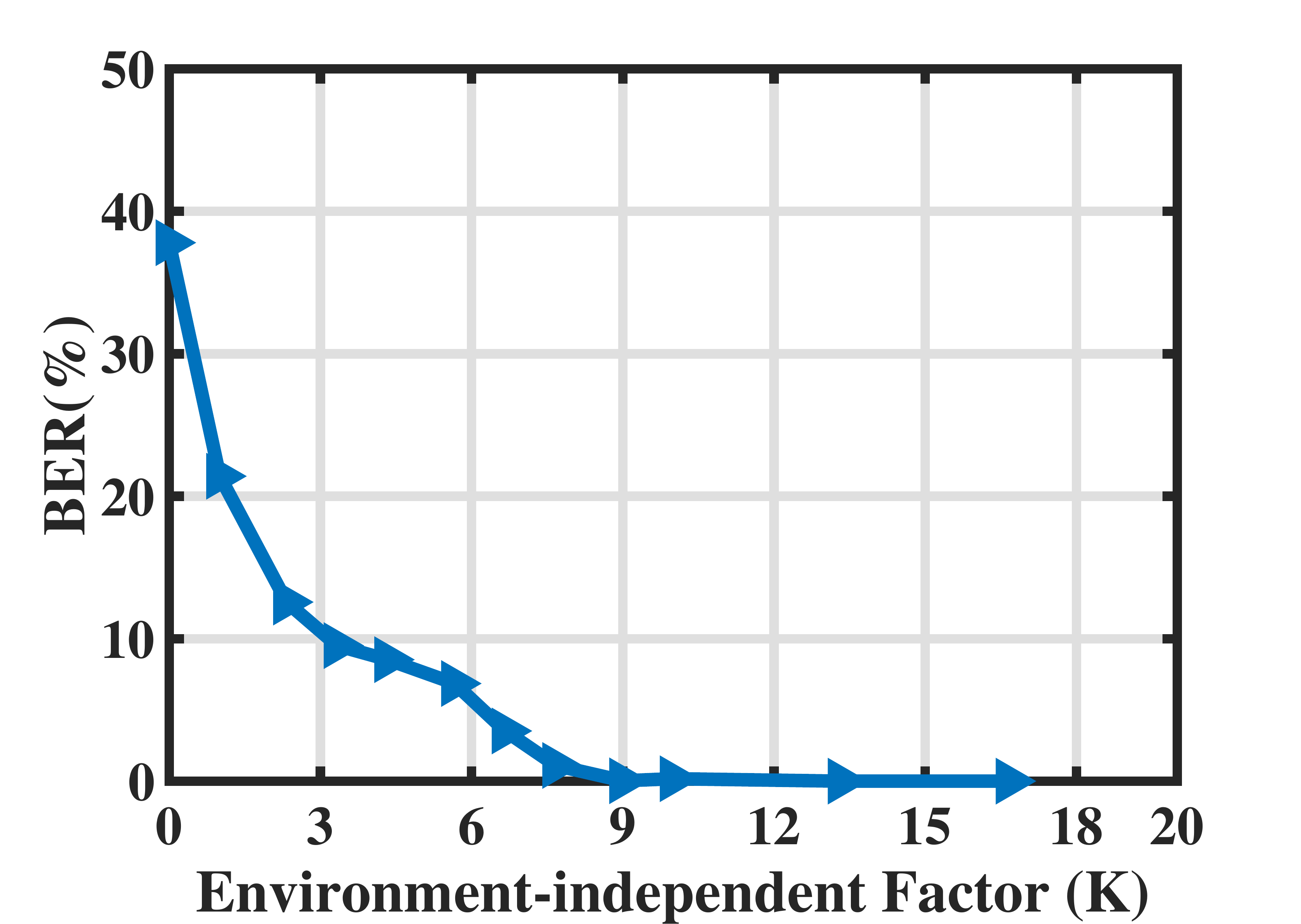}
	\label{FIG:impact_k}
    } 
    \subfloat[\footnotesize Different bitrates of ES-Tag distance]{
	\includegraphics[width=0.22\textwidth]{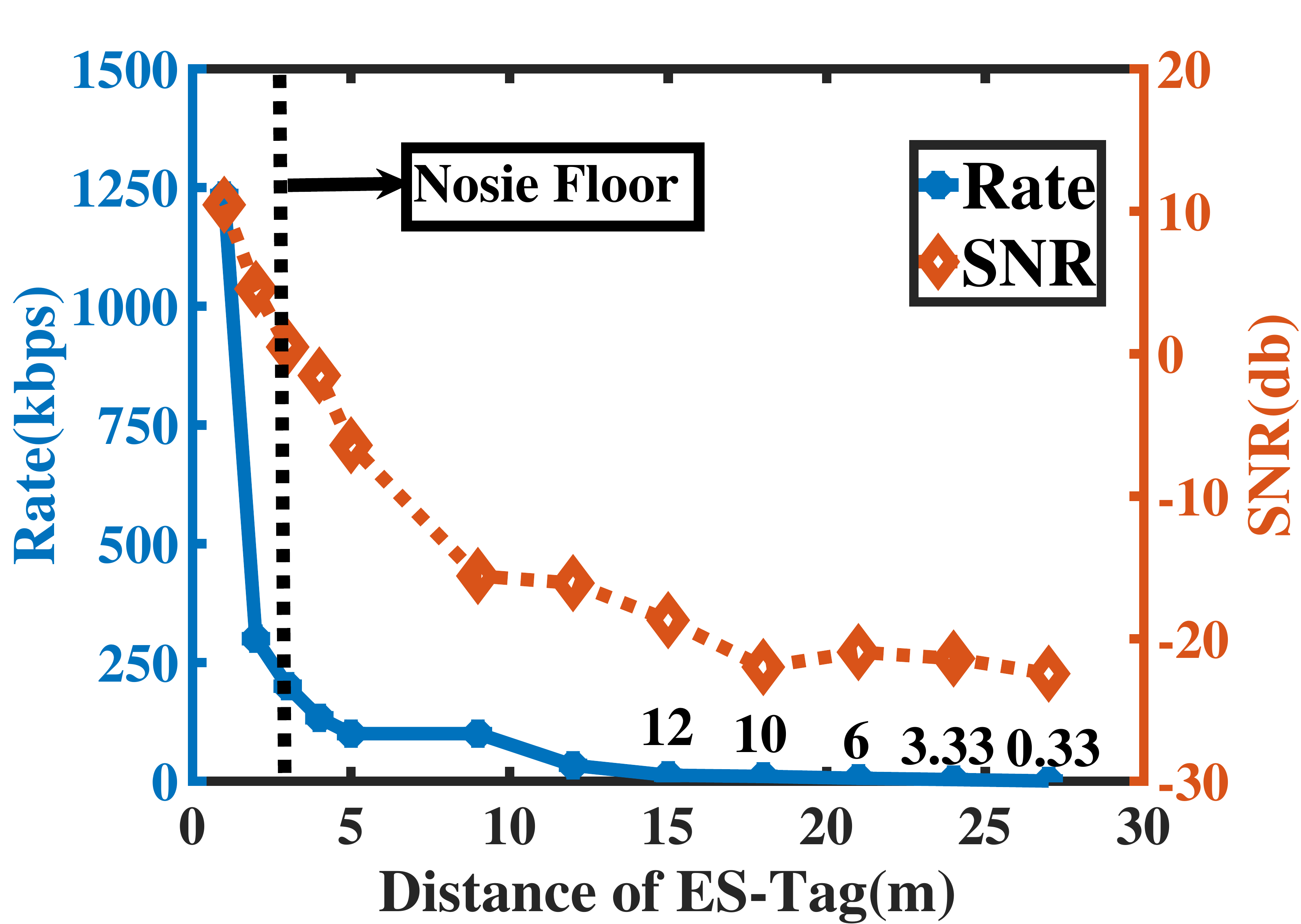}
	\label{FIG:rate_snr}
    } 
	\caption{Macro Benchmark}
	\label{FIG:macro_exp}
\vspace{-0.1in}
\end{figure}

\subsection{Macro Benchmark}
\label{sec:macro}

 \textbf{Estimation process of $K$:} 
 The $K$ estimation is very important to ADR in our system.
 According to Eq.~\ref{EQN:rate}, we could preset the value of $K$ which guarantees the fastest bitrate under the same SNR and bandwidth for one tag. 
 In these experiments, we estimate $K$ from $0$ to $16$ step by 1 which include 17 integers. We conduct the experiments for 50 times on 10 random location each in the entire room and calculate the mean BER. Fig.~\ref{FIG:impact_k} shows the relationship between the mean BER and $K$. We observe that the mean BER drops with the K increasing. When BER below $1\%$, $K$ can be estimated as $8$. 

 \textbf{Bitrates of Distance:} After estimating the value of $K$, we test the performance of \systemname in an playground. The results are as show in Fig.~\ref{FIG:rate_snr}. The solid line is the result of bitrate adaptation with the distance increases. The dotted line with star point represents the change in SNR with distance. The dotted line shows that the signal is below the noise when the distance is greater than $3.5m$. The max communication distance is expanded to $27m$, and bit rate adaptation range is from $0.33kpbs$ to $1.2Mbps$. As the distance increases, the bitrate and SNR will decrease to varying degrees. We can find a blank area between the solid line and the dotted line with star point, since we sacrificed a part of the bitrate in order to ensure communication stability.

\subsection{Micro Benchmark}
\label{sec:micro} 
 \textbf{Impact of Chirp-BW:} In order to investigate the impact of chirp signal bandwidth (BW) on \systemname, we select 5 different BW from $100K$ to $500K$. The distance between ES and tag is set as $5$m. The sampling rate of the ES is set to 10Mhz. Each experiment is repeated for 100 times, where the average results as well as the standard errors are fully evaluated. 
 
\begin{figure}[!ht]
	\centering
	\vspace{-0.15in}
    \subfloat[\footnotesize Impact of Chirp-BW]{
	\includegraphics[width=0.24\textwidth]{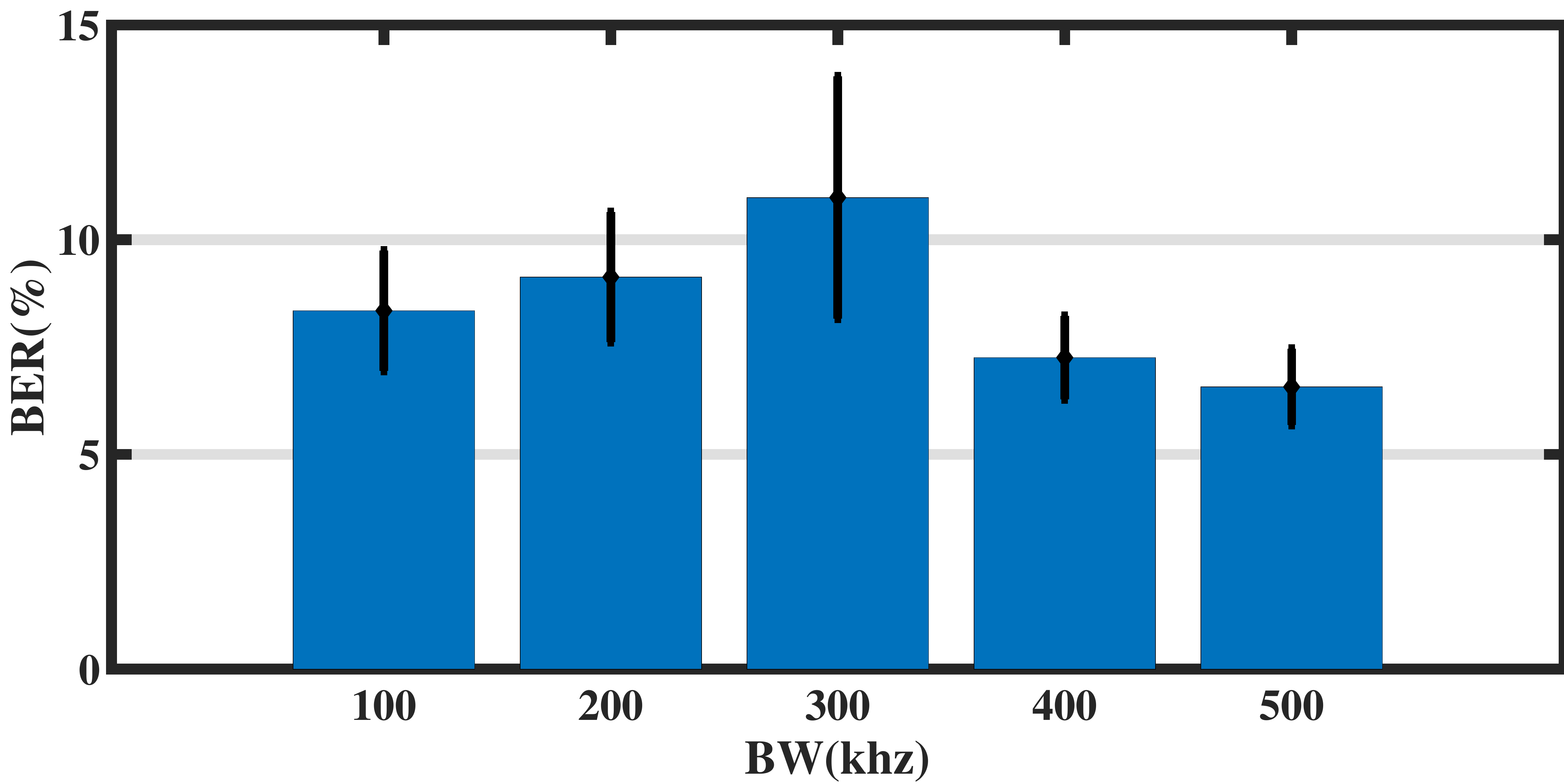}
	\label{FIG:impact_BW}
    }
    \subfloat[\footnotesize Impact of sampling rate]{
	\includegraphics[width=0.24\textwidth]{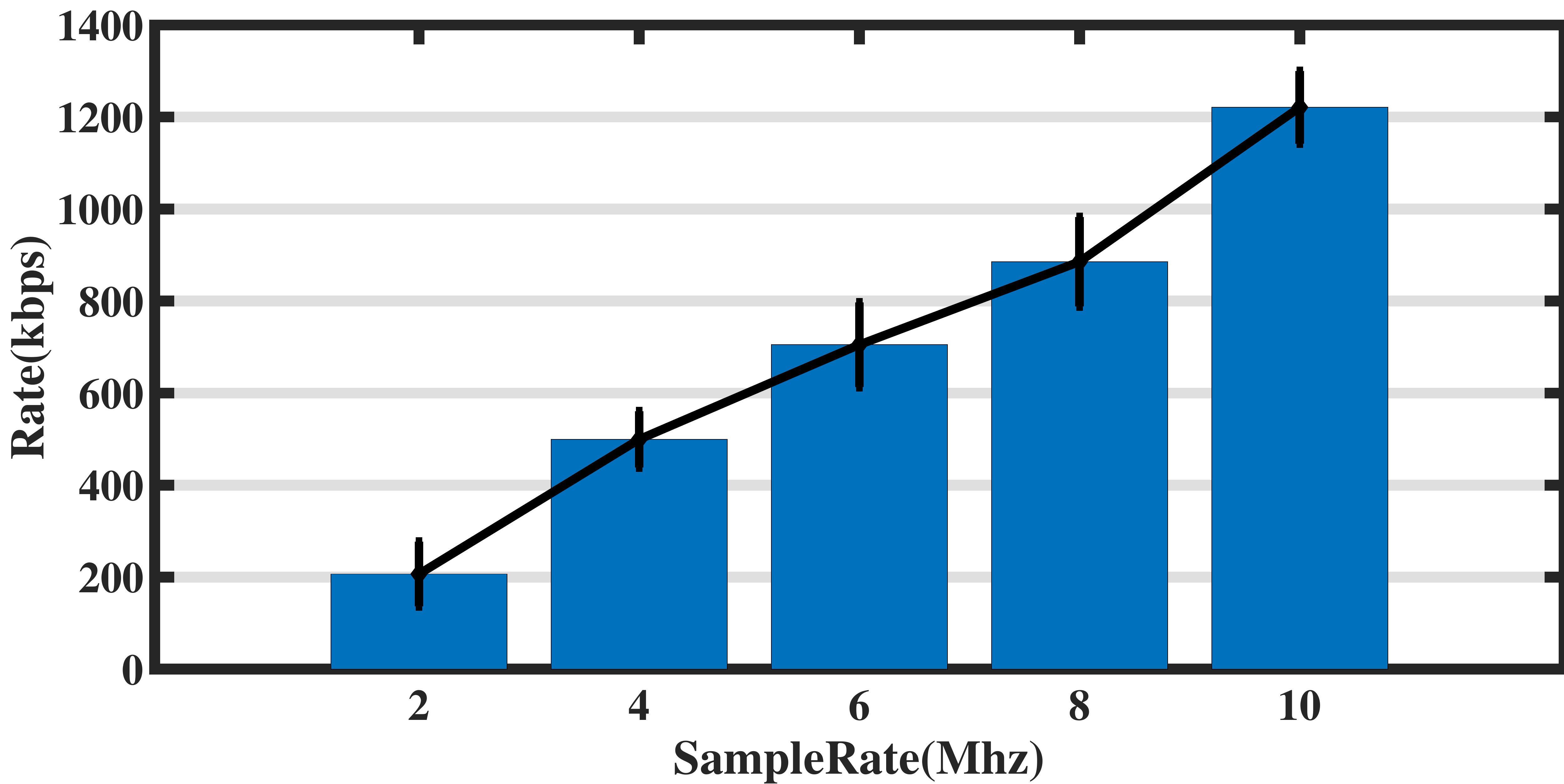}
	\label{FIG:impact_sample}
    } 
	\caption{Illustration of optimizing parameters}
	\label{FIG:micro_exp}
	\vspace{-0.2in}
\end{figure}

 We noticed an interesting phenomenon, as shown in Fig.~\ref{FIG:impact_BW}, the BER of the maximum bitrate at different bandwidths is basically same. As stated in Sec.~\ref{sec:chirp-ook}, $ Peak \propto N $, and the value of $Peak$ represents the ability to demodulate. So the BER of the maximum bitrate is just related to the number of sampling points, not the bandwidth of the chirp signal. Changing the bandwidth of the chirp does not bring gain to the BER, when the sampling rate is constant.
 
  \textbf{Impact of Sampling Rate:} In order to verify that the sampling rate of the ES is the main influence factor of system performance, we selected 10 different sampling rates from $2M$ to $10M$ for experiments. The distance between ES and tag is set as $5m$. The bandwidth of the chirp is set as $500K$. From Fig.~\ref{FIG:impact_sample}, we can see bitrate increases with the sampling rate increase.

\section{Conclusion} 
\label{sec:concl}
This paper has introduced a system of self-reliant bitrate adaptation without feedback for backscatter communication. Base on the channel symmetry, the backscatter tags can change the on-of key unit length for adaptive bitrate by detecting the chirp signal strength from the ES. And system can work under the noise floor.
In this system, the ES-Tag distance in communication was expanded to $27m$, and the supported bitrate adaptation range was $0.33kbps$-$1.2Mbps$. At the receiver, we designed a dynamic threshold algorithm for decoding, and the threshold has short-term memory to prevent errors caused by mutation. The future work will start with how to quickly and accurately detect the signal strength of the ES.
\vspace{-0.05in}

{ 
\bibliographystyle{abbrv}
\bibliography{COOK}
}

\end{document}